\documentclass[letterpaper]{article}

\usepackage[T1]{fontenc}

\usepackage{geometry}
\usepackage{setspace}

\usepackage{achemso}
\setkeys{acs}{articletitle = true}

\usepackage{graphicx}
\usepackage{float}
\newfloat{scheme}{htbp}{los}
\floatname{scheme}{Scheme}
\floatname{chart}{Chart}
\newfloat{graph}{htbp}{loh}

\usepackage{chemformula} 
\usepackage[version = 4]{mhchem} 

\usepackage{booktabs}
\usepackage{bm}
\usepackage{url}
\usepackage{array} 
\usepackage{longtable} 
\usepackage{hyphenat}
\usepackage{placeins}
\usepackage{amsmath}
\usepackage{amssymb}
\usepackage{bm}
\usepackage{comment}
\usepackage{wrapfig}

\usepackage{authblk}

\author[1]{Vid Ravnik}
\author[1,2,3]{Urban Bren}
\author[4]{Tine Curk*}
\affil[1]{Faculty of Chemistry and Chemical Engineering, University of Maribor, Smetanova ulica 17, 2000 Maribor, Slovenia}
\affil[2]{The Faculty of Mathematics, Natural Sciences and Information Technologies, University of Primorska, Glagoljaška ulica 8, 6000 Koper, Slovenia}
\affil[3]{Institute for Environmental Protection and Sensors, Beloruska ulica 7, 2000 Maribor, Slovenia}
\affil[4]{Department of Materials Science and Engineering, Johns Hopkins University, 3400 North Charles Street, Baltimore, Maryland 21218, United States}

\title{Designing Multivalent Copolymers for Selective Targeting of Multicomponent Surfaces}

\date{*Email: tcurk@jhu.edu}

\begin{document}

\maketitle

\begin{center}
\textbf{Abstract}
\end{center}

\small

\begin{wrapfigure}{r}{0.54\textwidth}
    \centering
    \vspace{-10pt}
    \includegraphics[width=0.52\textwidth]{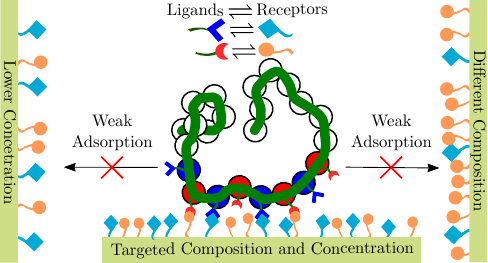}
    \vspace{-10pt}
\end{wrapfigure}

Selective targeting of membranes with a specific receptor profile is an ongoing challenge in targeted drug delivery. We investigate the adsorption of copolymers on a multicomponent receptor-covered surface using grand-canonical Monte Carlo simulations and demonstrate that polymers can be designed to target a particular receptor density profile. To achieve this, the ligand profile on the polymers should match the targeted receptor profile, and the ligand--receptor affinity should be inversely proportional to the ligand profile. While the same can be obtained using multivalent nanoparticles, the entropic effects due to polymer conformations significantly enhance the binding selectivity of multivalent polymers compared to nanoparticles. Surprisingly, the ligand distribution on the polymer plays a crucial role, whereas the persistence length does not. The optimal selectivity to the overall receptor concentration is obtained by the Poisson distribution of ligands (random copolymer), whereas the maximal selectivity to a specific receptor profile is obtained by a defined sequence of grouped alternating ligands (regular copolymer). Interestingly, the regular copolymer can become anti-selective when ligands of the same type are in homogenous blocks, showing that specific ligand distribution qualitatively affects the targeting ability.
These findings suggest that sequence control is necessary to selectively target a specific density profile of membrane receptors using linear copolymers. 

\normalsize
\vspace*{1em}


\section{Introduction}

The field of nanomedicines strives to  apply nano-sized tools to the prevention and treatment of disease~\cite{duncan2011nanomedicine}. It is concerned with designing drug delivery systems which preferentially deliver drugs to desired tissues. One of the main applications of nanotherapeutics is cancer treatment, some strategies make use of passive targeting of tumors using the EPR effect~\cite{matsumura1986new, maeda2000tumor}, while others try to overcome some of the limitation of passive targeting by actively targeting disease-specific biomarkers~\cite{cho2008therapeutic,bertrand2014cancer,pearce2019insights}. 
Selectively targeting cancer cells can be especially difficult, since they often do not express a unique biomarker, but can instead only be recognized through the overexpression of certain markers also present in healthy cells. In such a system, selective targeting of the malignant cells is crucial for avoiding damage to healthy cells and reducing the negative side effects of a given treatment~\cite{shadidi2003selective,shi2017cancer,liyanage2019nanoparticle}.

 The selectivity of binding is far from the only consideration when designing nanotherapeutics~\cite{blanco2015principles, duncan2017polymer,zhao2020targeting}, but it is an important one. 
In this work we focus on a particular strategy that has proven effective at increasing the selectivity of binding to a surface overexpresing certain biomarkes, namely multivalent binding~\cite{krishnamurthy2006multivalency,chittasupho2012multivalent,varner2015recent}.
Multivalent binders are species (nanoparticles, polymers) that interact simultaneously with a multitude of binding units. We henceforth refer to the moieties carried by the multivalent binders as ligands, and the moieties present on a (cell) surface which we are trying to target as receptors, whether or not these are ligand and receptors in the biological sense.
The combinatorial entropy of multivalent interactions leads to superselectivity, a sharp, faster-than-linear increase in the number of bound multivalent probes with the density of receptors~\cite{martinez2011designing}. Thus, multivalency facilitates selective targeting of surfaces on the basis of the density of receptors or markers.

Multivalent binding is commonly found in nature. Living organisms often rely on superselective multivalent interactions between biomolecular complexes and cells, in protein–glycan interactions on the cell surface~\cite{collins2004cell}, immune recognition~\cite{carlson2007selective}, intercellular transport~\cite{zahn2016physical,hoogenboom2021physics}, adhesion, and host recognition by bacteria or viruses~\cite{overeem2021dynamic,overeem2021multivalent}, as well as in many other examples~\cite{mammen1998polyvalent,dubacheva2023determinants}.
Many types of multivalent construct have been reported in the scientific literature~\cite{dubacheva2023determinants}, ranging from polymers~\cite{dubacheva2015designing}, proteins~\cite{richter2005kinetics,curk2022controlling}, antibodies~\cite{nores1987density,carlson2007selective}, viruses~\cite{overeem2021dynamic} to nanoparticles~\cite{hong2007binding,scheepers2020multivalent,linne2021direct}.
In addition to targeting a single overexpressed receptor on the cell surface, theoretical~\cite{curk2017optimal,caplan2005targeting} as well as experimental~\cite{robinson2008targeting,Tian_2020_prec_nmed} studies have shown promising results when using nanoparticles to target a profile of multiple receptors. Interactions with multiple receptor types facilitate targeting of a membrane that does not possess a single unique overexpressed receptor or marker~\cite{curk2017optimal}.

A prominent branch of nanomedicine development are polymer therapeutics, which present an alternative to nanoparticle based approaches~\cite{duncan2011polymer,connor2017polymers,atkinson2018polymer,ashford2021highway}.
Multivalent polymers can also exhibit higher selectivities compared to nanoparticles due to stronger cooperative effects of polymer binding to a surface~\cite{dubacheva2015designing,dubacheva2023determinants}.
We are specifically referring to chelate cooperativity, which arises in multivalent entities from the interactions between surface receptors and multiple ligands on the same entity, combined with allosteric cooperativity where binding of some ligands facilitates binding of other ligands~\cite{Hunter2009_coop, Ercolani2011_coop}. Stronger correlations between ligand binding within the same entity indeed suggest that copolymers could be more selective than nanoparticles in targeting a specific receptor concentration profile; where a specific receptor profile is defined by the membrane composition, i.e., the membrane concentrations of all relevant receptors. 
Moreover, previous studies have investigated the adsorption of linear polymers with a variety of different distributions of surface-binding functional groups on the polymer backbone, which revealed that the distribution of ligands exerts a significant effect on polymer adsorption~\cite{marques1989block, balazs1990models, balazs1991effect, balazs1998stabilizing, eskilsson1998interfacial, jhon2009effect}.
However, the potential and ability of multivalent copolymers to target a specific receptor profile in multicomponent surfaces currently remain unknown. This study aims to fill that gap in our understanding by investigating the optimal design of multivalent copolymers for selective targeting.

\begin{figure}[htb!]
    \centering
    \includegraphics[width=14cm]{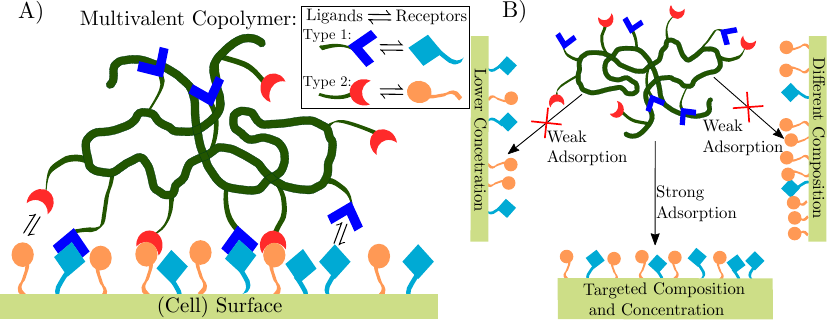}
    \caption{Surface targeting with copolymers. 
    A) Schematic representation of a multivalent copolymer containing two types of ligands (colored blue and red) binding to a surface containing two types of receptors (colored light blue and light red). 
    B) The goal of selective targeting: the probe binds strongly to a surface with the targeted receptor profile and concentration; while binding weakly/not binding to surfaces with a different profile or lower concentration of receptors. 
    } 
    \label{fig:multival_schema}
\end{figure}

We focus on a system of linear multivalent copolymers with two or three distinct types of ligands that adsorb to a planar surface containing different types of receptors. The ligands carried by polymers can bind to the corresponding receptors on the surface, schematically depicted in Figure \ref{fig:multival_schema}A. 
We strive to make our results applicable to a range of multivalent selective targeting applications, such as e.g. host--guest chemistry~\cite{dubacheva2014superselective,dubacheva2015designing}, biorecognition~\cite{wittmann2013bridging,ponader2014carbohydrate,rodriguez2015design}, or targeting overexpressed biomarkers. We therefore apply a scale-free model where a polymer is represented as a chain of course-grained polymer beads~\cite{pierleoni2007soft}. Each bead represent a polymer coil whose radius of gyration is similar to the surface receptor size. 
We investigate how the ligand distribution, interaction strength, and backbone rigidity affect the selectivity using grand-canonical Monte Carlo simulations. The goal is to design copolymers that selectively target a specific composition and concentration of receptors, as illustrated in Figure~\ref{fig:multival_schema}B.

\section{Model and Methods}

\subsection{Simulation model}

We employ a coarse-grained model and study multivalent copolymer adsorption on a multicomponent membrane using grand-canonical Monte Carlo. 
To keep the model as general as possible, we employ a soft-blob model where each polymer consists of $N_\textnormal{b}$ beads connected by harmonic springs. Each bead represents a flexible polymer coil its interactions are given by soft Gaussian repulsion~\cite{pierleoni2007soft} (soft-blob model). The soft-blob model has been previously utilized to study the adsorption of multivalent polymers~\cite{dubacheva2014superselective,dubacheva2015designing} or polymer-coated nanoparticles~\cite{curk2017optimal,curk2020first}. The main advantage of the soft-blob model are transferable potentials, we can represent a particular polymer with different numbers of beads (i.e., many small beads or fewer larger ones) without charging the interaction potentials. Each polymer bead in our model represents a polymer coil whose radius of gyration is similar to the size of the surface receptors. 
When studying the effect of polymer persistence length, we use the standard bead-spring model where each bead represents an individual monomer described by WCA repulsion and a three-body angular term that controls polymer stiffness~\cite{tschop1998simulation} (bead-spring model). We apply the soft-blob model for a flexible polymer because it is computationally faster than the bead-spring model due to soft Gaussian potentials compared to hard-core WCA. However, we find that both models yield qualitatively the same results with a simple rescaling of interaction by $0.1$~$k_\textnormal{B}T$ (see SI, Fig. S2).

The membrane is modeled as a flat surface with area $S$ that contains receptors of multiple different types. The membrane is described by a receptor composition vector $\bm{N_{\textnormal{r}}} = \big[N_{\textnormal{r,1}}, ..., N_{\textnormal{r,n}} \big]$, where $N_{\textnormal{r,i}}$ represents the number of receptors of type $i$ while $n$ is the total number of receptor types. The total number of receptors is $N_{\textnormal{rec}} = \sum^n_{i=1} N_{\textnormal{r,i}}$. The receptor composition $\bm c$ determines the receptor densities as a fraction of the total and is defined as $\bm c =   \big[ c_1, ..., c_n \big] = \bm{N_{\textnormal{r}}} / N_{\textnormal{rec}}$. The receptors are mobile within the membrane and interact via a short-range hard-sphere potential between each other and via bonding with ligands in polymers. 

To model a multivalent copolymer, specific beads contain side-chain ligands (Fig.~\ref{fig:multival_schema}). Beads carrying ligands can bind to one surface receptor but are otherwise described by the same interactions as beads without ligands. A polymer contains $\bm{N_{\textnormal{l}}} = \big[N_{\textnormal{l,1}}, ..., N_{\textnormal{l,n}} \big] $ ligands where $N_{\textnormal{l,i}}$ represents the number of ligands of type $i$ and each bead can contain at most one ligand.
The total number of ligands on a polymer is $N_{\textnormal{lig}} = \sum^n_{i=1} N_{\textnormal{l,i}}$,
and the corresponding copolymer ligand composition~$\bm l$ is defined in the same way as the receptor composition, $\bm l =  \big[ l_1, ..., l_n \big] = \bm{N_{\textnormal{l}}} / N_{\textnormal{lig}}$. 
Ligands can form valence-limited bonds with receptors, a ligand can bind to at most one receptor at a time, and \emph{vice versa}. The bond interaction is described with a symmetric interaction matrix $\bm \epsilon$ of size $n$ where each element $\epsilon_{{i,j}}$ denotes the binding free energy between ligand type $i$ and receptor type $j$.  $\epsilon_{{i,j}}$ represents the strength of the individual ligand--receptor interaction and is connected with the experimentally relevant dissociation constant for the particular interaction, $\beta \epsilon_{i,j}   \propto \ln(K_{\textnormal{d},ij})$, with $K_{\textnormal{d},ij}$ the ligand--receptor dissociation constant. Ligands can only bind to cognate receptors, that is, we do not consider cross-binding, $\epsilon_{i \neq j} = \infty$.
A flexible side chain linker is implicitly modeled with a weak harmonic potential between bound ligands and receptors.
We expect our findings are robust and model-independent, which is supported by the observation that bead-spring and soft-blob models yield consistent results (see Supporting Information for additional details about the model).

We performed grand-canonical Monte Carlo (GCMC) simulations of multivalent polymer adsorption onto a surface covered with receptors. Polymer insertion/deletion was handled using the Rosenbluth Configuration-Bias Monte Carlo method~\cite{frenkel2002book}. 
From GCMC simulations we calculate the average number of bound polymers~$\langle N_\textnormal{bp} \rangle$ and the fraction of the surface occupied by polymers~$\theta$. We define a polymer ``footprint'' size as $a = V_\textnormal{poly}^{1/3}$, where $V_\textnormal{poly} = \frac{4\pi}{3} R_\textnormal{g}^3$ is the volume occupied by one polymer in the bulk solution at dilute conditions, with $R_\textnormal{g}$ being the radius of gyration of the polymer; $R_\textnormal{g} \approx R_\textnormal{b} N_\textnormal{b}^{\nu}$ for the soft-blob model with $R_\textnormal{b}$ the radius of gyration for an individual bead and the scaling exponent $\nu=0.588$ for good solvent conditions. The surface coverage is then  $\theta = \frac{\langle N_\textnormal{bp} \rangle}{S}a^2$, with $S$ as the area of the receptor covered surface
\footnote{We note that when studying the effect of polymer rigidity we still use the flexible case  $R_\textnormal{g}$ to determine the polymer "footprint" so that $\theta$ is always the same function of the number of bound polymers. For rigid polymers, the calculation of the surface coverage is not accurate, but a more accurate approach would only rescale $\theta$ values linearly and thus not affect selectivity, which is the main focus of our study.}.

We initially focus on a symmetric case targeting a surface with two distinct receptor types with equal numbers, $\bm{c^*} = [0.5, 0.5]$, where $\bm c^*$  is the targeted receptor composition. Interaction strengths are also symmetric, $\epsilon_{1,1} = \epsilon_{2,2} = \epsilon$.
The receptor surface density~$\sigma_\textnormal{R}$ is set to 50 receptors per flexible polymer ``footprint'' area, $\sigma_\textnormal{R} = 50/a^2$ and polymers consist of $N_\textnormal{b} = 48$ beads. We set the chemical potential of the polymers such that the bulk polymer concentration $\rho = 0.01/V_\textnormal{poly}$. 
The simulation box is $L_x=L_y = 100 R_\textnormal{b}\approx 10R_\textnormal{g}$ in lateral directions and $L_z =  60R_\textnormal{b}\approx 6R_\textnormal{g}$ perpendicular to the surface.
We performed grand canonical simulations to $10^9$ cycles (where a cycle represents $N_\textnormal{b}$ individual Monte Carlo moves of either individual bead displacement, receptor displacement or polymer insertion/deletion), with three repeats of each simulation to obtain an averaged result. 
To investigate polymer binding cooperativity, we performed free-energy calculations using the Wang-Landau method~\cite{wang2001efficient,landau2004new} and determine the dependence of the free energy on the number of bonds formed, $F(\lambda)$, and the receptor composition, $F(c_1)$.
Additional details about the implementation of the simulations can be found in the Supporting Information. 

\subsection{Quantifying selectivity}

\begin{figure}[htb!]
    \centering
    \includegraphics[width=14cm]{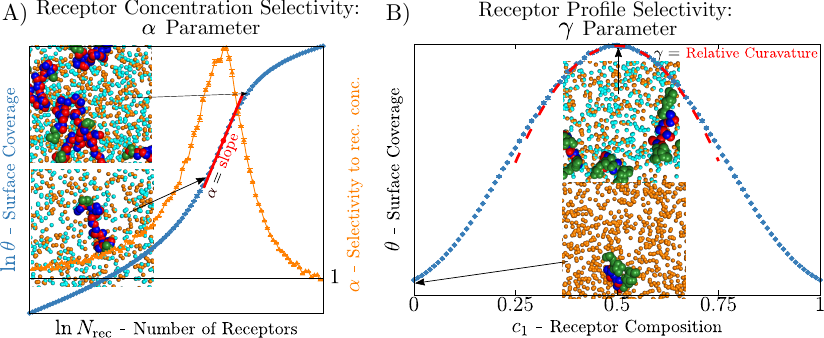}
   
    \caption{A schematic representation of the definition of the selectivity parameters $\alpha$ and $\gamma$. 
    A.) $\alpha$ represents selectivity to the number of receptors, it is calculated as the slope of a plot of $\ln\theta$ against $\ln N_{\textnormal{rec}}$, defined by Eq.~\eqref{eq:alpha}. 
    B.) $\gamma$ represents selectivity to the receptor profile by measuring the relative curvature of the surface coverage $\theta(\bm{c})$ with respect to the receptor composition at the targeted composition defined by Eq.~\eqref{eq:gamma}. The insets in A and B feature examples of equilibrium system configurations where the small orange and cyan spheres represent receptors of two different types. The large green, red, and blue spheres represent polymer beads, where green spheres carry no ligands, while red and blue beads carry ligands that bind with orange and cyan receptors, respectively. The polymers have $\bm l = [0.5,0.5]$ with central-alternating ligand distribution. The graph and snapshots in A) are calculated with $N_{\textnormal{lig}} = 40$ and $\epsilon = -2.8$, the graph and snapshots in B) with $N_{\textnormal{lig}} = 16$ and $\epsilon = -3.5$. 
    }
     
    \label{fig:sele_parameters}
\end{figure}

We are interested in two distinct types of selectivity, selectivity to a change in the total number of receptors on the surface and selectivity to a change in the receptor composition.
To quantify selectivity to the number of receptors we use the well-known selectivity parameter~ $\alpha$, first defined by Martinez-Veracoechea and Frenkel~\cite{martinez2011designing}:

\begin{equation} \label{eq:alpha}
        \alpha = \frac{d\ln{\theta}}{d\ln{N_{\textnormal{rec}}}} \;,
\end{equation}

where $\theta$ is the surface coverage, i.e., the fraction of the surface covered by the bound polymers, and $N_{\textnormal{rec}}$ is the total number of receptors. High values of $\alpha$ ($\alpha > 1$) define the superselectivity regime, where an increasing number of receptors results in a faster-than-linear increase in the number of bound polymers, see Fig.~\ref{fig:sele_parameters}A.

To quantify selectivity to the receptor composition we define a new selectivity parameter, $\gamma$, which represents the relative curvature in the surface coverage with respect to the receptor composition around the targeted value ($\bm{c} = \bm{c^*}$), calculated as:
\begin{equation} \label{eq:gamma}
    \gamma = \det\Bigg(  \frac{
    -\bm{\mathcal{H}}\big[\theta(\bm{c})\big]
    }{\theta(\bm{c})  }    \Bigg)_{\bm{c} = \bm{c^*}} \;,
\end{equation}
where $\bm{\mathcal{H}}\big[\theta(\bm{c})\big]$ is the Hessian matrix of second derivatives of the surface coverage $\theta$ with respect to the receptor composition $\bm{c}$. Thus, $\gamma$ is defined as the (negative) relative curvature of $\theta$ at the targeted composition $\bm c^*$. Note that for a two-component surface, $\gamma$ simplifies to a single term $\gamma=-1/\theta(c_1^*)\left(\frac{d^2\theta}{dc_1^2}\right)_{c_1=c_1^*}$; see also Figure \ref{fig:sele_parameters}B which illustrates the calculation of $\gamma$.
This definition [Eq.~\eqref{eq:gamma}] mirrors the selectivity~$\mathcal{S}$ based on the curvature of the binding free energy in Ref.~\citenum{curk2017optimal}. Basing the definition on the surface coverage~$\theta$ rather than free energies provides a more direct measure of the targeting selectivity and represents a direct counterpart to $\alpha$.

\newpage
Table \ref{tab:eps_Fb} provides a quick reference for parameters and variables commonly mentioned throughout this work.

\begin{table}[htb!]
\centering
\caption{Quick reference for common parameters and variables}
\label{tab:eps_Fb}
\begin{tabular}{l||l}
\hline
\hline
$\epsilon$,   $\epsilon_{i,j}$ & Binding strength parameter, represents the free energy  \\
 &  of individual receptor--ligand interactions, $\beta \epsilon_{i,j}   \propto \ln(K_{\textnormal{d},ij})$  \\
\hline
$N_\textnormal{lig}$ &  Number of ligands on polymers, up to 1 ligand per polymer bead  \\
\hline
$l_1$, $c_1$  & Fractions of ligands and receptors of type 1\\
\hline
$k_\phi$ & Parameter describing the strength of the angular potential, \\
& connected with persistence length $\ell_\textnormal{p}\approx \langle l\rangle \beta k_\phi$ for $k_\phi > k_\textnormal{B}T$, \\
\hline
$\theta$ & Surface coverage, the fraction of the surface covered by polymers \\
\hline
$\alpha$, $\gamma$ &  Selectivity to the receptor concentration and composition \\
& (see Eqs.~\eqref{eq:alpha} and \eqref{eq:gamma}, Fig.~\ref{fig:sele_parameters})\\
\hline
$F_\textnormal{b}$                          & Free energy of binding of polymers, $\beta F_\textnormal{b} \propto \ln (K_{\textnormal{d}, \textnormal{polymer}})$
\\
\hline
$f_\textnormal{b}$                          & Free energy of binding per ligand, $f_\textnormal{b} = F_\textnormal{b}/N_\textnormal{lig}$
\\
\hline
$F(\lambda)$, $F(c_1)$         & Dependence of the free energy on the number of formed bonds $\lambda$  \\
 & and the fraction of type 1 receptors $c_1$. \\
\end{tabular}
\end{table}

\section{Results and Discussion}

\subsection{Polymer design: Ligand Distribution}

This work aims to provide guidelines for designing multivalent polymers to enhance selective targeting. To this end, we studied the effect of the ligand distribution on the adsorption and selectivity of multivalent polymers.
Most theoretical models of multivalent interactions~\cite{martinez2011designing,dubacheva2014superselective,dubacheva2015designing,curk2017optimal} employ a mean-field approximation where ligands positions are not explicitly considered.
However, the position of a ligand in the polymer chain can significantly influence the properties of the polymer. For example, a ligand on the edge of a polymer likely has a lower entropic penalty for binding than a ligand in the middle of the polymer. More importantly, multiple ligands adjacent to each other in the chain contour can bind to a surface as a ``train'' with a much lower entropic penalty per ligand, compared to ligands far apart that form ``loops''. 

\begin{figure}[htb!]
    \centering
    \includegraphics[width=12cm]{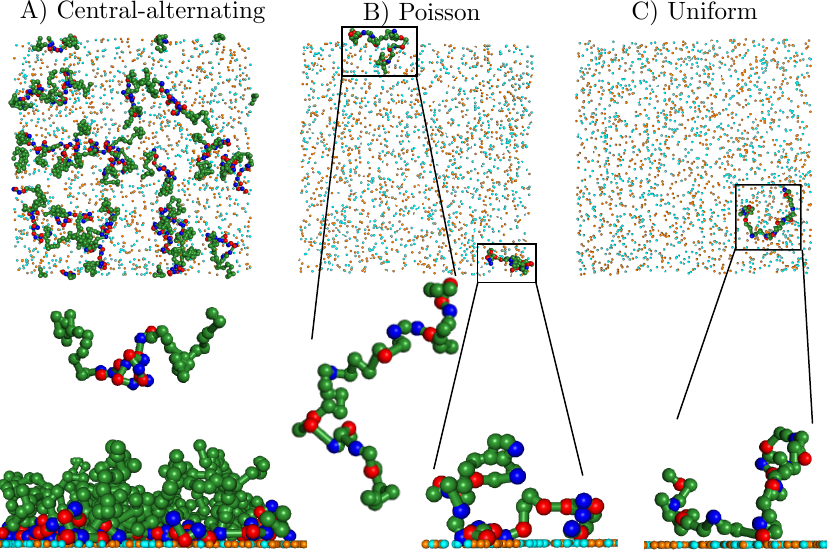}
   
    \caption{ Examples of simulation snapshots for different ligand distributions with the same binding strength and number of ligands.
    A,B,C) Snapshots for central-alternating, Poisson, and uniform copolymers, respectively. Top view on the top, close-ups of the side view below.  Green spheres depict polymer beads not carrying ligands, while red and blue spheres represent beads with ligands of two different types. Smaller orange and cyan spheres denote surface receptors that can bind with red and blue polymer beads, respectively. 
  Parameters:  symmetric ligand and receptor profile $\bm{l} = \bm{c} = [0.5, 0.5]$,  $\beta\epsilon = -3.75$, $N_\textnormal{lig} = 16$. 
    } 
    \label{fig:poly_snaps}
\end{figure}

\begin{figure}[htb!]
    \centering
    \includegraphics[width=12cm]{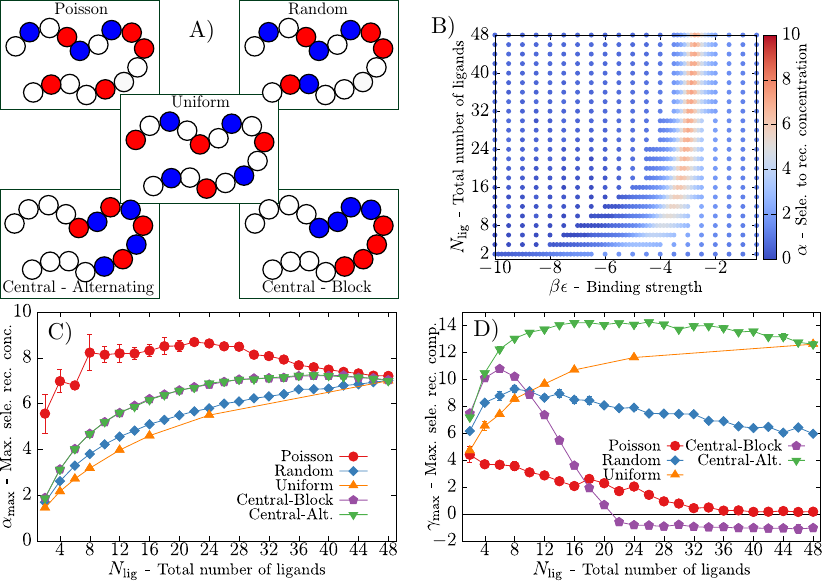}
   
    \caption{ The effect of the ligand distribution on selectivity:
    A) The five types of ligand distributions investigated in this work (see text for exact definitions).
    B) Heatmap of selectivity~$\alpha$ for different numbers of ligands $N_\textnormal{lig}$ and binding strengths $\epsilon$ for central-alternating copolymers. 
    C,D) The maximum value of selectivities (maximum over all $\epsilon$)  for different ligand distributions and numbers of ligands $N_\textnormal{lig}$.   
    Data obtained using a symmetric ligand profile $\bm{l} = [0.5, 0.5]$.  In most cases the uncertainties are smaller than the symbol size, lines are guides for the eye. 
    } 
    \label{fig:lig_dist_max_sele}
\end{figure}

Throughout this work, we considered five different ligand distributions on polymer chains, see Figs.~\ref{fig:poly_snaps} and~\ref{fig:lig_dist_max_sele}A for a schematic representation.
Through a typical synthesis process of a copolymer, randomness in the number and distribution of ligands can be expected. We capture this with polymers on which the number of ligands is Poisson distributed (with $N_\textnormal{lig}$ as the expected value) and the ligands are randomly distributed along the polymer (Poisson copolymers). Each polymer inserted into the simulations has a randomly determined number and distribution of ligands of each type, which means the ligand composition of individual polymers can also vary between polymers.
We also consider polymers with a constant number of ligands that are randomly distributed (Random copolymers), i.e. every polymer has exactly the same stoichiometry but the ligand positions are random. The remaining ligand distributions considered in this study represent well-defined regular copolymers: we studied copolymers with ligands distributed uniformly on the polymer backbone in a regular alternating pattern (Uniform copolymers) and central copolymers that have all of the beads with ligands in a block in the center of the polymer and all of the beads without ligands in tails on the edges. We studied two versions of central copolymers, a variant where the types of ligands in the central block alternate (Central-alternating copolymers) and a variation where the ligands of a particular type are all placed together in a block within the larger central block of all ligands (Central-block copolymers). 
We note that each bead in the simulations does not model a single monomer, but a short flexible polymer coil whose size is comparable to the size of receptors.  

Figure \ref{fig:poly_snaps} shows a comparison of equilibrium configurations between central-alternating, Poisson, and uniform copolymers with identical binding strength, ligand, and receptor composition.  In Fig.~\ref{fig:poly_snaps}A  we observe many central-alternating polymers, the ligand beads are bound to the surface, forming cooperative ``trains'', while the tails without ligands stick away from the surface. Poisson and uniform copolymers (Fig.~\ref{fig:poly_snaps}B,C) exhibit fewer polymers bound to the surface due to loops between the bound ligands. 
Configurations of central-block copolymers are qualitatively similar to central-alternating, while those of random copolymers resemble those of Poisson copolymers. This shows that ligand sequence can substantially affect binding. 

To investigate the selectivity of these five ligand distributions, we performed simulations with different ligand--receptor binding strengths~$\epsilon$ and numbers of ligands per polymer~$N_{\textnormal{lig}}$ and calculated the corresponding selectivity parameters $\alpha$ and $\gamma$.
Figure \ref{fig:lig_dist_max_sele}B shows the dependence of selectivity~$\alpha$ on $\epsilon$ and $N_{\textnormal{lig}}$ for central-alternating copolymers as an example. We see that the selectivity is low when the binding is weak, it increases and peaks with increasing binding strength $\epsilon$ and falls again at higher $\epsilon$ when the surface becomes saturated (cf.\ Fig.~\ref{fig:sele_parameters}A). The position of the maximum selectivity moves to less negative $\epsilon$ (weaker ligand--receptor bonds) at higher $N_\textnormal{lig}$ due to multiple simultaneous interactions~\cite{curk_2017_ch3}.

We calculated an analogous dataset for the other four ligand distributions for both selectivity parameters (Supporting Information, Figs. S6, S7) and found the position of the maximum is qualitatively similar in all cases, less cooperative ligand distributions move the position of the maximum to more negative binding strengths at low $N_\textnormal{lig}$ (SI, Fig. S5). 
The maximum value of the selectivity, however, can be widely different. We are interested in the maximum selectivities across all binding strength values $\epsilon$ for the different ligand distributions. To minimize statistical errors, the maximal selectivity at each value of $N_\textnormal{lig}$ was calculated by fitting a cubic spline to $\alpha(\epsilon)$ and $\gamma(\epsilon)$ within a range $\epsilon \pm 0.15$ from the highest value 
and finding the maximum (Fig.~\ref{fig:lig_dist_max_sele}C, D). Uncertainties in the maximal selectivity were estimated by calculating the standard deviations via the bootstrap method~\cite{efron1985bootstrap,johnson2001introduction}.  The uncertainties for Poisson copolymers are much greater due to the additional randomness in the structure of the polymers.

\noindent\textbf{Selectivity to the total receptor concentration }

Our results show that, surprisingly, Poisson-distributed ligands provide the highest selectivities to the overall receptor concentration~$\alpha$ (Fig.~\ref{fig:lig_dist_max_sele}C). This is counter-intuitive since the polydispersity in the distribution of binders usually flattens the adsorption or binding curve. Moreover, regular copolymers are generally more specific than random copolymers~\cite{jhon2009effect}. However, with a random Poisson distribution the polymers that preferentially adsorb are those with a higher-than-average number of ligands, which increases the selectivity. 
We did not find any previous specific mention of the increase in selectivity with the Poisson distribution of ligands, but this result is consistent with mean-field theoretical models using only a single type of ligand~\cite{curk_2017_ch3}.  Nevertheless, it is surprising that Poisson copolymers exhibit selectivity higher than that of the specifically designed regular copolymers. 

At a constant number of ligands (no Poisson distribution), central (alternating and block) copolymers display the highest selectivity~$\alpha$, followed by random copolymers, while uniform copolymers exhibit the lowest selectivity.  We propose that the reason for this lies in the cooperativity between the ligands. Central copolymers have ligands next to each other, which leads to cooperative behavior; when a ligand binds to the surface, it brings adjacent ligands close to the surface as well, making them more likely to bind together in "trains"~\cite{marques1989block, balazs1990models, balazs1991effect, balazs1998stabilizing, eskilsson1998interfacial, jhon2009effect} (cf.~Fig.~\ref{fig:poly_snaps}). 
Uniform copolymers, on the other hand, have ligands further apart, which decreases the cooperative effect. Moreover, the parts of the polymer between ligands in uniform copolymers crowd the surface, forming "loops", whereas in central copolymers the beads without ligands stick out as "tails" away from the surface with receptors. These effects lead uniform copolymers to exhibit more gradual adsorption and thus lower selectivity compared to central copolymers.
Similar results were found in a study on multivalent copolymers with a single type of ligand, where copolymers with all ligands in a block significantly outperformed uniform copolymers~\cite{tito2014optimizing}. As expected, random copolymers fall in between these two limiting cases. We find statistically no difference in the selectivity~$\alpha$ between central copolymers with alternating ligand types and with all ligands of a single type in a block, implying that the distribution of ligand types does not have a discernible effect on the selectivity to the number of receptors. 
At the maximal number of ligands ($N_\textnormal{lig} = 48 = N_\textnormal{b}$, all polymer beads carry ligands), the selectivity~$\alpha$ for all distributions converge. 
This is because uniform and central-alternating copolymers with $N_\textnormal{lig} = N_\textnormal{b}$ are identical, while central-block and random copolymers differ only in the distribution of ligand types.

\noindent\textbf{Selectivity to the receptor composition }

We now turn our attention to selectivity to changes in the receptor profile~$\gamma$ (cf.~Fig.~\ref{fig:sele_parameters}B).
Contrary to the case of $\alpha$ discussed above, Poisson-distributed ligands exhibit among the lowest $\gamma$ values (Fig.~\ref{fig:lig_dist_max_sele}D). 
It was argued before using an analytical model that for Poisson-distributed ligands on multivalent nanoparticles the binding free energy cannot exhibit a minimum at an arbitrarily chosen receptor composition~\cite{curk2017optimal}. Indeed, we find $\gamma$ values that are close to zero with Poisson copolymers
\footnote{The $\gamma$ results for Poisson copolymers are calculated using the curvature of $\theta$ from the full range $c_1 = 0$ to 1, different from the remaining ligand distributions which use a curvature by approximating a parabola within 0.3--0.7. This is due to large uncertainties in $\gamma$ for Possion copolymers when using the smaller range.}.

Central-alternating copolymers exhibit the highest $\gamma$ values for any number of ligands~$N_{\textnormal{lig}}$. Uniform copolymers exhibit significantly lower selectivity than central-alternating copolymers, we again attribute this difference to the decrease in cooperativity between ligand binding.
Moreover, contrary to $\alpha$, the $\gamma$ selectivity for central-block copolymers differs greatly from that of central-alternating copolymers. For $N_\textnormal{lig} = 2$ central-block and central-alternating copolymers are identical, therefore, their selectivities are the same. Surprisingly, as $N_\textnormal{lig}$ increases, the maximal selectivity of central-block copolymers decreases greatly compared to other ligand distributions and even turns negative. This means that for any value of binding strength $\epsilon$, central-block copolymers are \emph{anti}-selective, the surface coverage $\theta$ does not have a maximum, and in fact shows a minimum at the targeted receptor composition. Random copolymers can contain all random local arrangements of ligands and fall in the middle between the three regular cases. 
We note that even when each bead carries a ligand ($N_\textnormal{lig} = 48$) $\gamma$ for random copolymers is significantly lower than for central-alternating/uniform (which are identical at $N_\textnormal{lig} = 48$). This implies that a ligand distribution where ligand types are randomly distributed in a block in the center of the polymer (this would likely be much simpler to synthesize experimentally than alternating ligand types) would perform significantly worse than central-alternating, and that alternation of ligand types is indeed key to achieving high $\gamma$ selectivity.

These results show that the distribution of ligand types qualitatively affects the selectivity~$\gamma$. 
We propose that the reason for the importance of ligand type distribution lies in cooperative regions of the same ligand type. 
To provide an example, in the case of polymers with equal numbers of two types of ligands ($l_1=l_2 = 0.5$), the targeted surface for such polymers is one with an equal distribution of two types of receptors ($c_1^* = c_2^* = 0.5$). If we consider the adsorption of such polymers to a surface with only one type of receptor ($c_1 = 1$), the beads carrying ligands type 2 play no part in the binding, since we do not consider cross-binding $\epsilon_{12}=\infty$.
This means that central-alternating copolymers behave similarly to uniform copolymers with a single type of ligand (beads not involved in binding are located between beads with ligands) and therefore the cooperative binding effect is much weaker.
On the other hand, the relevant ligands in central-block copolymers still form a block, which results in strong cooperative adsorption as a train on the surface, even when the surface contains a single receptor type. The simulation results show that this effect can be very strong, to the point that the polymers become anti-selective. For sufficiently large blocks (large  $N_\textnormal{lig}$), the binding of a single block to a surface at $c_1 = 1$ is stronger than the binding of two blocks at lower receptor concentrations ($c_1 = c_2 = 0.5$) and thus $\gamma$ becomes negative.

Summarizing the effect of the ligand distribution, the random Poisson distribution provides the optimal selectivity to the receptor concentration~$\alpha$. Therefore, if we want to apply multivalent polymers to selective targeting of surfaces based on overall receptor concentration, the ligand distribution on the polymers does not need to be precisely controlled. 
On the other hand, when targeting a particular receptor profile, random fluctuations in the number of ligands and even the positions of the ligands negatively impact the selectivity~$\gamma$. Therefore, precise control of the polymer sequence is necessary to successfully target a specific receptor composition. Copolymers with alternating types of ligands in a block in the center of the polymer exhibit the highest selectivity~$\gamma$. Moreover, block copolymers exhibit very low selectivity and even become anti-selective, $\gamma<0$,  when the size of the block is large.
The exact position of the ligand block is not of great importance as shown in  Ref.~\citenum{tito2014optimizing} that studied polymers with a single type of ligand and demonstrated that polymers with the block in the center give slightly higher $\alpha$ selectivity than polymers with the block at the edge, but the difference was minimal.
Because central copolymers with alternating ligand types exhibit the highest selectivity to the receptor profile $\gamma$ and simultaneously very high selectivity to receptor concentration we will mainly focus on central-alterating copolymers throughout the rest of this work. 

We note that our study did not consider the effect of polydispersity in the polymer size, only in the number of ligands. We expect that size polydispersity would likely lead to a decrease in selectivity. 
Moreover, this study was performed in conditions where the basic requirements for superselectivity (a large number of possible bound microstates) are met for all ligand distributions (many receptors in range to bind). In conditions where this does not hold we expect deviation from our predictions. For example, if receptors are large and ligand linkers very short such that adjacent ligands cannot be bound simultaneously we expect that uniform copolymers, which have adjacent ligands separated further apart so that they can both bind to receptors simultaneously, could outperform central copolymers.
Since each soft-blob bead represents a flexible polymer coil, we emphasize that the results of our study do not show that adjacent polymer monomers should carry ligands for optimal selectivity. Instead, the distance between ligands should approximately match the minimum distance between receptors, so that neighboring ligands can bind.   
Within the studied conditions (an abundance of receptors) receptor mobility does not have a large effect on surface coverage and selectivity (SI Fig. S3). When studying systems with fewer receptors their mobility could start to play a role in selectivity as polymers recruit nearby receptors. Previous studies have shown that receptor mobility can enhance ($\alpha$) selectivity and shift the selectivity maximum to a lower number of receptors~\cite{dubacheva2019multivalent,morzy2022significance}. In the cases where receptors are in fixed positions, large inhomogeneities in those positions could also affect results, with recent work showing that uniformity in the receptor distribution can enhance selectivity~\cite{di2020recruitment,xia2023role}. 

\subsection{Polymer Rigidity}

In addition to ligand distribution, polymer bending rigidity represents yet another tunable parameter that distinguishes polymers from nanoparticles and other multivalent constructs. To model semiflexible polymers we used the bead-spring model~\cite{tschop1998simulation}, which represents each monomer by a single bead and features a harmonic bond between consecutive beads, a Weeks-Chandler-Andersen (WCA) repulsive interaction between the beads, and an additional three-body angular potential term. The angle potential~$ U_{\textnormal{angle}}$ controls the rigidity of the polymer,
\begin{equation} 
        U_{\textnormal{angle}} = k_\phi (1-\cos\phi) \;,
        \label{eq:angle_pot}
\end{equation}
where $\phi$ is the angle between the vectors of two consecutive bonds, and the prefactor $k_\phi$ determines the polymer stiffness. The persistence length is $\ell_\textnormal{p}\approx \langle l\rangle \beta k_\phi$ for $k_\phi > k_\textnormal{B}T$, where $\langle l\rangle$ is the average bond length and $\beta = 1/(k_\textnormal{B}T)$, while for $k_\phi \ll k_\textnormal{B}T$ we recover the flexible polymer with $\ell_\textnormal{p}\approx \langle l\rangle/2$ since $\langle l\rangle$ represents the single Kuhn length. 

The flexible polymer case ($k_\phi = 0$) yields quantitatively the same results for selectivity as the soft-blob model used throughout the rest of this work.  A comparison and additional details about the polymer models can be found in the Supporting Information (Fig. S2). This indicates that the design rules for optimal selectivity (Fig.~\ref{fig:lig_dist_max_sele}) do not depend on the system size, and apply both to a small length-scale, e.g. biorecognition where specific polymer monomers bind to specific sites on a target protein, or a large-length scale where a long ligand-functionalized polymer binds to multiple different proteins or receptors.

\begin{figure}[htb!]
    \centering

    \includegraphics[width=12cm]{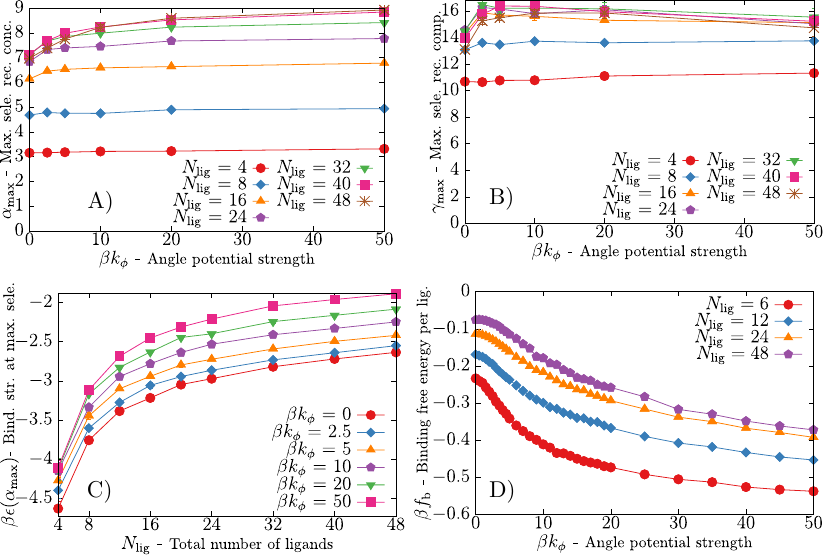}
   
    \caption{The effects of polymer rigidity on selectivity.
    A,B) The dependence of the maximum selectivities $\alpha$ and $\gamma$, on the rigidity of the polymer for different numbers of ligands in the polymer. $k_\phi$ is the strength of the angular potential which controls polymer stiffness [Eq.~\eqref{eq:angle_pot}].
    C) The dependence of the binding strength $\epsilon$  at maximal $\alpha$.
    D) The dependence of the free energy of binding per ligand, $f_\textnormal{b}$, on $k_\phi$ at $\beta \epsilon = -2.5$. Parameters: $\bm{l} = [0.5, 0.5]$. Uncertainties are smaller than symbol size.
    } 
    \label{fig:max_sele_cen_eps_b}
\end{figure}

We studied the effect of polymer stiffness on selectivity by calculating the dependence of the maximal selectivity $\alpha$ and $\gamma$ on the strength of the angular interactions $k_\phi$. Maximal selectivity was calculated by optimizing the receptor--ligand binding strength $\epsilon$ as described for data in Figs. \ref{fig:lig_dist_max_sele}C,D.
We focused on central copolymers with alternating ligands since this type of copolymer has proven optimal for $\gamma$ while still exhibiting high $\alpha$ values. The results for maximal $\alpha$ and $\gamma$ as a function of angle potential strength (persistence length) are depicted in Fig.~\ref{fig:max_sele_cen_eps_b}. The studied range of $k_\phi$ is large, from flexible polymers ($k_\phi=0$) to stiff, rod-like polymers ($\beta k_\phi = 50$). We find that, surprisingly, changing the rigidity does not have a strong effect on either $\alpha$ or $\gamma$.
Increasing the rigidity of the polymer increases $\alpha$ at high $N_\textnormal{lig}$ by up to 30\%. However, in all other cases, changing the persistence length by nearly two orders of magnitude has no significant effect on maximal selectivities.

We interpret that the reason for the small effect of persistence length is that increasing the polymer rigidity introduces competing effects on polymer selectivity. 
First, it increases the effective size of the polymers by increasing the radius of gyration $R_\textnormal{g}$ (see SI, Fig. S1), which in turn increases the size of the site required for a polymer to adsorb. Thus the effective activity of the polymers is increased, which results in a decrease in selectivity~\cite{dubacheva2019multivalent} (see SI, Fig. S4), as seen in Figure \ref{fig:max_sele_cen_eps_b}D. On the other hand, increasing polymer stiffness increases cooperative binding by increasing the propensity to form "trains" while decreasing the number of "loops" formed when polymers adsorb to a surface~\cite{hsu2013effect,linse2010polymer}. The increased number of "trains" suggests more cooperative binding, which we have shown increases the selectivity of the adsorption. Thus, the two effects of increased $R_\textnormal{g}$ and increased cooperatively have opposite effects on selectivity, explaining why changing the persistence length altogether does not exert a large effect on the maximal selectivity.

We demonstrate the increase in binding cooperativity by showing that 
the optimal binding strength curve shifts to less negative $\epsilon$ at higher $k_\phi$, indicating that polymer binding is stronger at higher persistence lengths (Fig.~\ref{fig:max_sele_cen_eps_b}C). 
This is confirmed directly by calculating the free energy of binding per ligand~$f_\textnormal{b}$ using the Wang-Landau method~\cite{wang2001efficient}
and biasing the simulations in the number of formed bonds, see Figure \ref{fig:max_sele_cen_eps_b}D.
We observe the binding energy decreasing with increasing $k_\phi$ for different values of $N_\textnormal{lig}$, further demonstrating the proposed increased cooperativity in ligand binding with polymer rigidity. 

We note that polymer persistence length may have a more significant impact on selectivity in conditions where a change in rigidity could significantly alter how many receptors the ligands of a particular polymer can access. For example, if the spatial distance between neighboring ligands on a flexible polymer chain is smaller than the receptor size, the increase in persistence length would increase the spatial distance between ligands allowing them to bind more receptors. Moreover, if the maximal number of receptors that can fit into a polymer footprint is smaller than the number of ligands, increasing the persistence length would increase the polymer footprint allowing binding to more receptors and thus increasing the selectivity.


\subsection{Comparison with analytical theory}

The majority of theoretical models describing multivalent binding postulate the assumption that the ligand binding is uncorrelated~\cite{martinez2011designing,dubacheva2014superselective,dubacheva2015designing,curk2017optimal,curk2020first}.  This means that ligands bind independently with no allosteric cooperativity; the binding of one ligand has no effect on the binding of the remaining ligands (apart from occupying a receptor). 
To investigate whether these models apply to multivalent copolymers, we compare our simulation results with the theoretical model for multicomponent multivalent binders~\cite{curk2017optimal}. This model assumes that different ligands bind independently and the free energy of binding per ligand~$f_\textnormal{b}$ can be calculated analytically from the ligand and receptor profiles and the binding strengths,
\begin{equation} \label{eq:theory_fb}
    f_\textnormal{b}(\bm{l},\bm{N_r},\bm{\epsilon}) = - \sum_i l_i \ln\left(1+\sum_j c_j N_\textnormal{rp} e^{-\beta \epsilon_{i,j}} \right) \;.
\end{equation}
where $N_\textnormal{rp}$ is the total number of receptors a polymer can bind with ($N_\textnormal{rp} = \sigma_\textnormal{r}a^2$). 
We can predict the surface coverage~$\theta$ from the binding free energy using the Langmuir isotherm,
\begin{equation} \label{eq:langmuir}
    \theta = \frac{z e^{-\beta F_\textnormal{b} }}{1+z e^{-\beta F_\textnormal{b} }} \;,
\end{equation}
where $z$ is the polymer activity and $F_\textnormal{b} = N_\textnormal{lig} f_\textnormal{b}$.  

We have shown that selectivity can be strongly increased by strengthening the correlations between ligand binding (Fig.~\ref{fig:lig_dist_max_sele}). However, at this point, it is unclear if the additional correlations affect the general design rules for selective targeting of specific receptor composition profiles that were previously developed for nanoparticles~\cite{curk2017optimal}. The two design rules are:
\begin{enumerate}
\item The ligand profile should match the receptor profile it targets, $\bm l = \bm c^*$. 
\item Binding strength should be inversely proportional to targeted receptor density, $e^{-\beta \epsilon_{i,i}}  \propto  1/c^*_i$, and no cross-binding $e^{-\beta \epsilon_{i\ne j}} = 0$.
\end{enumerate}
This implies that every ligand has the same probability of being bound. The third design rule states that this probability should be $\approx 70\%$  (i.e., free-energy per bond is $f_\textnormal{b}=1.256k_\textnormal{B}T$) at optimal selectivity~\cite{curk2017optimal}.

\begin{figure}[htb!]
    \centering

    \includegraphics[width=12cm]{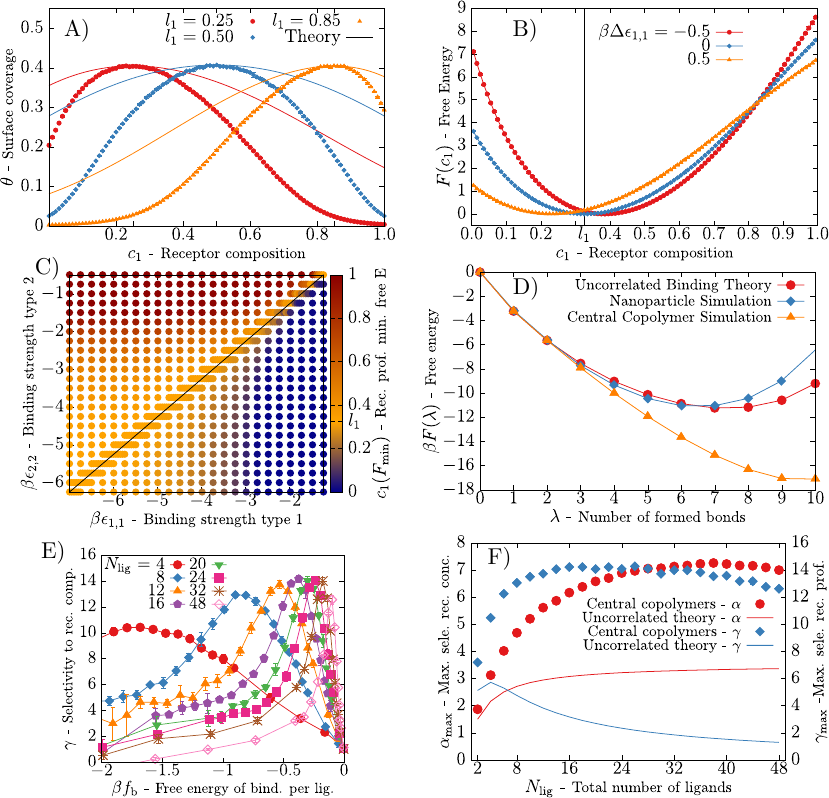}

    \caption{Design rules for copolymers, and comparison with nanoparticle theory (Eq.~\eqref{eq:theory_fb}).
    A) Surface coverage~$\theta$ at different ligand profiles comparing polymer simulations (symbols) and theory (Eqs.~\eqref{eq:theory_fb},\eqref{eq:langmuir}) (solid lines) at the same bulk polymer concentration and maximum value of surface coverage, and following the first two design rules. 
    B)  Dependence of the free energy on the receptor composition $F(c_1)$. Copolymers have $\bm l = [0.325,0.675]$, $\beta\epsilon_{2,2} = -3$ and $\Delta \epsilon_{1,1}$ represents the deviation of $\epsilon_{1,1}$  from the second design rule.
    C) The targeted composition; receptor composition~$c_1$ at the minimum free energy $c_1(F_\textnormal{min})$, for a spectrum of binding strength values at $\bm l = [0.325,0.675]$. The black line follows the theoretical design rule, $\beta\epsilon_{2,2} = \beta\epsilon_{1,1} + \ln\left( l_2/l_1 \right)$, at which the targeted receptor profile indeed matches the ligand profile, $c_1 = l_1$ (orange color). 
    D) Dependence of the free energy on the number of formed bonds at $N_\textnormal{lig}=10$ for analytical theory (red circles) and simulation data for nanoparticles (blue diamonds; taken from Ref.~\citenum{curk2017optimal}) and central-alternating copolymers (orange triangles).
    E) The dependence of selectivity $\gamma$ on the free energy of binding per ligand $f_\textnormal{b}$. 
    F) A comparison of the maximal selectivity for polymers (symbols, cf. Fig.~\ref{fig:lig_dist_max_sele}C and D) and the theoretical model (lines). 
    Parameters: All polymer simulations use central-alternating copolymers with $N_\textnormal{lig}=20$, 40, 40, 10  for the A,B,C,D panels, respectively.
    } 
    \label{fig:np_comp}
\end{figure}

We investigate to what extent these design rules apply to linear polymers. 
We find that the maximum in surface coverage $\theta$ occurs exactly at $c_1 = l_1$  (Fig.~\ref{fig:np_comp}A), which is in perfect agreement with the first theoretical design rule on the position of the maximum. However, there is a significant disagreement between the uncorrelated ligands theory and polymer simulations in the curvature of $\theta(c_1)$, indicating larger $\gamma$ selectivity of polymers, which we will discuss below.
Polymer simulations are also in agreement with the second design rule; any deviation from the second design rule causes the polymer to miss-target the surface composition (Fig.~\ref{fig:np_comp}B).
Furthermore, we calculate the targeted receptor composition (i.e, the composition at which the binding free energy is minimized) over a full spectrum of interaction strengths $\epsilon_{1,1}$ and $\epsilon_{2,2}$, and find that it indeed matches the second design rule (Fig.~\ref{fig:np_comp}C). 
Interestingly, at stronger binding, we observe a larger area of $\epsilon_{1,1}$, $\epsilon_{2,2}$ values that target the correct receptor composition. While this extra leeway in polymer design appears very useful, we note that selectivity generally decreases at very negative binding strengths.

The preceding figures have shown that copolymer design for composition targeting should follow the first two theoretical design rules. 
This is surprising given the strong cooperative effects in polymer binding which are absent from analytical theory.  The cooperative effects are shown by comparing the dependence of free energy on the number of formed bonds~$F(\lambda)$ for polymers, nanoparticles, and analytical theory (Fig.~\ref{fig:np_comp}D). Nanoparticle behavior is relatively well described by the assumption of uncorrelated ligand binding, with noticeable deviations only at very large numbers of bonds~\cite{curk2017optimal}.
However, central copolymers differ greatly from the uncorrelated ligand theory, with significantly lower values of $F(\lambda)$ at a higher number of bonds, which confirms our earlier claim of strong cooperative binding in polymers due to the formation of "trains". 

We further investigate the effect of cooperative binding by calculating the dependence of selectivity~$\gamma$ on the free energy of binding per ligand~$f_\textnormal{b}$ (Fig.~\ref{fig:np_comp}E).
The dependence of $f_\textnormal{b}$ on $\epsilon$ for different $N_\textnormal{lig}$ was calculated via Wang-Landau simulations and combined with the data for the dependence of $\gamma$ on $\epsilon$ and $N_\textnormal{lig}$ (cf.\ Fig~\ref{fig:lig_dist_max_sele}D) to plot  $\gamma$ against $f_\textnormal{b}$. We note that because $f_\textnormal{b} =1/N_\textnormal{lig} \sum_\lambda e^{-\beta F(\lambda)}$ is calculated from the dependence of free energy on the number of formed bonds $F(\lambda)$, the value of $f_\textnormal{b}$ depends on the height constraint of the polymer which influences the free-energy to form the first bond $F(1)$. Following previous nanoparticle analysis~\cite{curk2017optimal}, we chose the height constraint such that the first two bonds follow the uncorrelated ligands theory (this also applies to other $F(\lambda)$ calculations through this work, Fig.~\ref{fig:np_comp}D and~\ref{fig:max_sele_cen_eps_b}D). 
We find the $\gamma$ maximum at $f_\textnormal{b}\approx -0.4k_\textnormal{B}T$, which is about three times smaller in magnitude than the theoretical prediction [third design rule: $\beta  f_\textnormal{b} \approx - 1.256$], which we attribute to the cooperative binding of the polymers [Fig.~\ref{fig:np_comp}D]. Thus, due to correlations in ligand binding, polymers exhibit much higher maximal selectivities than those predicted by uncorrelated binding theory for nanoparticles (Fig.~\ref{fig:np_comp}F).

\begin{figure}[htb!]
    \centering
    \includegraphics[width=7cm]{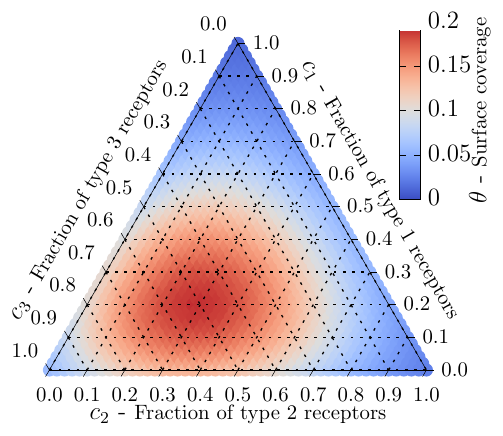}
    \caption{A heatmap of the surface coverage dependence of $\theta$ on the distribution of receptors $\bm{c} = \big[c_1, c_2, c_3\big]$ with 3 types of ligands and receptors. The polymers had randomly distributed ligands with $\bm{N_{\textnormal{l}}} = \big[4,6,10\big]$ ($\bm{l} = \big[0.2,0.3,0.5\big]$). The binding strength followed the first two design rules ($\epsilon_{i,i}\propto \ln(l_i)$) with $\beta \epsilon_{3,3}=-4.0$.
     } 
    \label{fig:3_lig_rec_types}
\end{figure}

Herein lies the main message of this work. Multivalent copolymers exhibit selectivities higher than those of nanoparticles and offer significant opportunities to improve selective targeting by influencing polymer-specific design parameters, such as the distribution of ligands. 
Another parameter that can significantly improve selective targeting is the number of distinct receptor and ligand types. Figure~\ref{fig:3_lig_rec_types} displays an example of targeting a surface with (random) copolymers containing three different types of ligands, where the binding strengths follow the first two design rules. The highest surface coverage arises at $\bm c^*\approx \bm l$, demonstrating that polymers indeed target the desired composition of the three receptor types.

\subsection{Design rules for targeting}

Here we summarize the design rules for targeting multicomponent surfaces using multivalent binders, both the results of this work and previous studies~\cite{curk2017optimal}. 
The following design rules apply to receptor composition targeting ($\gamma$ selectivity):

\begin{enumerate}
    
    \item  $\bm l = \bm c^*$: The ligand profile should match the target receptor profile.  \label{dr:c_l}
    \item$e^{-\beta \epsilon_{i,i}}  \propto  1/c^*_i$  (or equivalently $K_{\textnormal{d},ii}  \propto  c^*_i$): The ligand--receptor binding strength should be inversely proportional to targeted receptor density, cross-binding should be minimized $e^{-\beta \epsilon_{i\ne j}} = 0$. \label{dr:eps_c}
    \item Ligand binding should be weak, "each ligand independently should have the probability of being bound around 30-70\% depending on the degree of allosteric cooperativity between ligand binding. \label{dr:weak}
    \item Optimally targeting multicomponent surfaces requires precise control over the ligand distribution, ligands of different types should alternate in a regular pattern.  \label{dr:no_ran}
     \item Design for allosteric cooperativity when possible. For polymers this means that ligands should be close to each other on the chain contour, while still being sufficiently apart to allow binding with adjacent receptors. \label{dr:coop}
    \item Increasing the number of distinct ligand receptor types increases the potential selectivity. \label{dr:types}
\end{enumerate}

In practice we might want to maximize both $\alpha$ and $\gamma$ selectivity. Design rules \ref{dr:c_l} and~\ref{dr:eps_c} apply only to $\gamma$ selectivity and ensure that the maximal binding occurs at the desired receptor density profile. Weak ligand--receptor binding (design rule \ref{dr:weak}) applies to both $\alpha$ and $\gamma$, $\alpha$ is maximized at slightly weaker binding than $\gamma$, but the difference is small (see SI Fig. S5). 
The requirement of precise control over the ligand distribution (design rule \ref{dr:no_ran}) applies only to $\gamma$, whereas control over ligand distribution is not necessary to achieve high $\alpha$, in fact, we find that $\alpha$ is maximized in random (Poisson distributed) copolymers. Increasing the cooperativity of ligand binding (design rule \ref{dr:coop}) increases both $\alpha$ and $\gamma$. 
The effect of increasing the number of distinct ligand and receptor types (design rule \ref{dr:types}) on $\alpha$ has not been studied, theory (Eqs.~\eqref{eq:theory_fb} and~\eqref{eq:langmuir}) suggests that changing the number of distinct ligand and receptor types does not appreciably affect $\alpha$.

\section{Conclusions}

In summary, we investigated the optimal design of copolymers for selective targeting by performing grand-canonical Monte Carlo simulations. By considering five distinct ligand distributions, we show that the ligand sequence plays a crucial role. Surprisingly, selectivity to the overall receptor concentration~$\alpha$ is maximized by using random copolymers with a Poisson distribution of ligands per polymer, instead of regular copolymers with a defined sequence. Conversely, copolymers with a random distribution of the number of ligands per polymer are unable to target a specific receptor composition profile. The maximum selectivity to a receptor profile~$\gamma$ is obtained by regular copolymers with centrally located ligands.
We show that entropic effects and cooperativity between ligands are major factors in selective targeting. 
For example, polymers where all the ligands in a block in the center exhibit higher selectivities than polymers where ligands are distributed uniformly over the whole length of the polymer. Such central polymers exhibit a very strong cooperative effect due to the ability to bind as a "train". 
These findings support results previously obtained for the $\alpha$ selectivity of polymers with a single type of ligands~\cite{tito2014optimizing}. 
This study was performed at a constant concentration of receptors on the surface and a constant number of course-grained beads in polymers. We did not study in detail the relationship between the receptor separation (concentration and size) and the optimal ligand separation.  We expect that as receptor separation increases the optimal distance between ligands also increases, but the exact relation is currently unknown and ripe for further study. 

Interestingly, ligands of different types should alternate and not appear in a large block of, e.g., type 1 ligands followed by a large block of type 2 ligands, which would likely be much easier to synthesize experimentally. Alternation of ligands improves selectivity because it decreases cooperativity when binding to a surface where the receptor composition does not match the ligand composition on the polymer. Contrasting this to copolymers with blocks of the same type of ligands in the center, high cooperativity between the ligands when interacting with any surface can lead to drastically decreased selectivity. In fact, we show that central copolymers with ligand types in blocks can even be anti-selective; the surface coverage has a minimum at the targeted composition, showing that ligand distribution qualitatively affects the targeting ability.

We studied the effect of polymer rigidity on selectivity and found that changing the persistence length by two orders of magnitude exerts only a minor effect on maximal selectivities. Increasing the rigidity of polymers can increase the $\alpha$ selectivity by up to $30\%$ compared to flexible polymers, while it has no noticeable effect on the $\gamma$ selectivity. We explain this by the opposite effects of increased radius of gyration and of increased cooperative binding as trains upon increasing the rigidity.
Therefore, we conclude that the effect of backbone rigidity on selectivity 
is not substantial and copolymers will behave in a similar fashion regardless of the stiffness.

Comparing the simulation results with the theoretical design rules describing how to target a particular receptor composition, we find excellent agreement with our polymer simulations. Interestingly, the selectivities~$\alpha$ and $\gamma$ of polymers are much larger than predicted by the theory due to the entropic cooperative effects of the polymer ligand binding. Moreover, the location of the maximal selectivity for polymers occurs at weaker interactions (free energy per bond of only $f_\textnormal{b} \approx -0.4 k_{B}T$) compared to the analytical prediction, which we again attribute to the correlated (cooperative) ligand binding. 
These findings should be broadly useful for applications of specific targeting and binding in supramolecular chemistry and biology. By using multivalent copolymers, it is possible to improve selective targeting of cell surfaces and thus minimize the side effects of treatments with targeted drug delivery.

\section*{Acknowledgements}

The authors gratefully acknowledge the financial support of the Slovenian Research and Innovation Agency (ARIS) through program and project grants P2-0046, P1-0403, J1-2471, J1-1715, L2-3175, P2-0438, J1-4398, L2-4430, J3-4498, J7-4638, J1-4414, and J3-4497. The authors also thank HPC Vega for providing the computational resources necessary for this study. T.C. acknowledges support from startup funds provided by the Whiting School of Engineering at JHU.

\section*{Supporting information}

The following file is available free of charge.
\begin{itemize}
  \item Supporting Information (.pdf): Additional results and details about the model and methods.
\end{itemize}
\bibliography{refs.bib}

\end{document}